\documentclass{llncs}
\usepackage{hyperref}
\usepackage{booktabs}
\usepackage{amsfonts}
\usepackage{booktabs}
\usepackage{siunitx}
\usepackage{multirow} 
\begin{document}
\title{Machine Learning simulates Agent-based Model}
\titlerunning{Machine Learning}  
\author{Bernardo Alves Furtado\inst{1,2}}
\authorrunning{Furtado, B.} 
\tocauthor{Bernardo Alves Furtado}
\institute{Institute for Applied Economic Research (IPEA), Brazil\\
\email{bernardo.furtado at ipea.gov.br},\\ 
\and
National Council of Research (CNPq)}
\maketitle              
\begin{abstract}
Running agent-based models (ABMs) is a burdensome computational task, specially so when considering the flexibility ABMs intrinsically provide. This paper uses a bundle of model configuration parameters along with obtained results from a validated ABM to train some Machine Learning methods for socioeconomic optimal cases. A larger space of possible parameters and combinations of parameters are then used as input to predict optimal cases and confirm parameters calibration. Analysis of the parameters of the optimal cases are then compared to the baseline model. This exploratory initial exercise confirms the adequacy of most of the parameters and rules and suggests changing of directions to two parameters. Additionally, it helps highlight metropolitan regions of higher quality of life. Better understanding of ABM mechanisms and parameters' influence may nudge policy-making slightly closer to optimal level.
\end{abstract}

\section{Problem context}
Agent-based models have gained recognition since the original ideas of Schelling \cite{schelling_models_1969} and the seminal book of Epstein and Axtell \cite{epstein_growing_1996}. In fact, a number of books have extensively opened up the debate of applications on geography \cite{heppenstall2011agent}, economics \cite{hamill_agent-based_2016}, social sciences \cite{helbing_social_2012,johnson_non-equilibrium_2017} and policy \cite{geyer_handbook_2015}.

Much of the appreciation of ABM comes from its flexibility and high-level case abstraction. However, such adaptability comes with its great weaknesses, that is lack of benchmarking and reproducibility. The alternatives in the space of possibilities is so large that most authors keep expanding their own specific hands-on problem with little attention to comparability. Some efforts have been made towards resolving it \cite{grimm_standard_2006,grimm_odd_2010}.

Nonetheless, a single ABM can easily run into hundreds of thousands possible combinations of parameters, rules and behaviors. Sensitivity analysis driven by the modelers knowledge of the model and input data is usually applied to each parameter or rule, one at time, \textit{ceteris paribus}. Clearly, this is not an optimal solution. Consider further that, even a model that runs relatively fast, will demand some effort to run a high number of times. Calibration is thus much dependent on the modeler knowledge and control of the object and the ABM.

Given such context, the goal of this paper is to use input from an ABM in the form of configuration parameters and results (sample of 232) to train a set of Machine Learning algorithms having an optimal socioeconomic output as a target. The trained ML algorithms are then used to predict optimal outputs from  a random set of parameters (100,000) within a broader number of combinations. The analysis of actual and predict parameters may hint on optimal parameters and enable further understanding of the model. In practice, the ML algorithm simulates the running of the ABM for 100,000 times within seconds.

Besides this introduction, next section makes a brief presentation of the literature. Section 3 explains the methods and step-by-step procedures used and describes the data. Results and considerations conclude the paper.

\section{Machine Learning}

Machine Learning has been used together with ABM to automatic calibrate time discrete behavior of agents \cite{torrens_building_2011}, as suggested on an early framework by Rand \cite{rand2006machine}. Although there has not been extensive work on associating ML to ABM, the development of ML models and methods \textit{per se} are well established \cite{hastie_elements_2009,james_introduction_2015,tan_introduction_2006}

We exploratorily use four models of classification: 
\begin{itemize}
\item Random Forest Classification is a "perturb-and-combine technique[s]\footnote{Breiman, 'Arcing Classifiers', Annals of Statistics 1998.} specifically designed for trees. This means a diverse set of classifiers is created by introducing randomness in the classifier construction. The prediction of the ensemble is given as the averaged prediction of the individual classifiers". \href{http://scikit-learn.org/stable/modules/ensemble.html#forest}{(from sklearn documentation)} In the application, we use 10,000 forests using bootstrap, with a maximum depth of 15.
\item Support Vector Classification is a "support vector machine constructs a hyper-plane or set of hyper-planes in a high or infinite dimensional space”. The goal of the algorithm is to get the “hyper-plane that has the largest distance to the nearest training data points of any class" \href{http://scikit-learn.org/stable/modules/svm.html#svm-mathematical-formulation}{(from sklearn documentation)}. We use the kernel 'poly' with degree equals to 3. 
\item Neural Network. Within neural networks we use the Multi-layer Perceptron classifier (MPL) which is a model that "optimizes the log-loss function using LBFGS", i.e. an “optimizer in the family of quasi-Newton methods". \href{http://scikit-learn.org/stable/modules/generated/sklearn.neural_network.MLPClassifier.html}{from sklearn documentation} Activation method is `than`:  the hyperbolic tan function and maximum iteration is restricted to 2,000. 
\item and a 'Voting System' which averages the probabilities of the other classifiers in order to make a prediction. We used the `soft` voting option.
\end{itemize}

\section{Model and data}

We use data from a Spatially-Economic Agent-based Lab (SEAL) \cite{furtado_policyspace:_2018}, that focus on modeling fiscal analysis on municipalities based on three markets: goods, labor and real estate. SEAL uses official input data to enable an empirical model of 46 Brazilian metropolitan regions represented by their ACPs, i.e., their Areas of Concentrated Population. SEAL is an ABM that focus on relatively understanding the mechanisms and influences of parameters and rules on a number of global and local results that include production, unemployment, Quality of Life or commuting demand. Much of the appeal of SEAL is exactly the ability to change parameters and rules and observe the results. 

SEAL usually runs with a 2\% sample of the population for 20 years, saving and plotting monthly results. A typical run of that size takes between 5 and 60 minutes. Validation has been achieved \cite{furtado_policyspace:_2018} by comparing fiscal data and its distributions. Robustness was evaluated analyzing parameters, rules, and ACPs' sensitivity. 

All the tests, however, amount for a total of 232 runs of the model \footnote{Some runs included more than a single run so that stochasticity can be evaluated. Results, however, are recorded as averages. Even when there is more than a single run, the configuration parameters for a given result does not change. Thus, to be precise, we have results for 232 exclusive parameters configurations}. That includes an average of 7 intervals per parameter, two for boolean parameters (rules) and sometimes just a single run for a specific ACP. In total, we have 67 parameters, being 46 of those ACPs, three boolean, 5 that are freely in the interval $[0, 1]$. Finally, the other parameters have smaller reasonable interval of variance.

It is easy to see that no amount of feasible runs of SEAL would account for a systematic review of the space of possibilities. 

\subsection{Operational details and proposal}

This subsection details the step-by-step used in the analysis. Full code is available on \href{https://github.com/BAFurtado/MLsimulatesABM}{https://github.com/BAFurtado/MLsimulatesABM}.\footnote{Find also additional results and statistics at the repository.}

The first step is to organize the data. Every simulation of SEAL generates a configuration file in JSON format in a base directory. Inside such directory, each individual run is stored in folders that are numbered. Those folders keep the results of individual simulations. Additionally, every base directory saves an average results file, containing macro results of all runs for that given configuration of parameters. Thus, the reading of the data implies walking over all the folders containing parameters and results and organizing them into $X$ and $Y$ matrices. 

Such summary also applies specific rules to construct the target. We have developed the model so that the modeler may choose the parameters that configure the optimal social case. That includes choosing parameters and percentiles among the available options. We tested the combinations of higher Quality of Life Index (QLI) and GDP index and lower unemployment and GINI index.

Our baseline case is composed of those results in which the QLI are within the \textbf{top} 35\% of all results and simultaneously on the \textbf{bottom} 35\% of unemployment. \footnote{Our sample is extremely small - due to the ABM generating process - which led us to choose a rather generous top and bottom percentiles.} Thus, our target is given by the cases in which high quality of life with low unemployment is achieved in the last month of the simulation.

The next step is to train the three selected classification methods (Random Forest, SVC and a neural network/MPL) along with a voting system. 

Having the machines trained on the simulation results, a larger, random set of parameters is generated so that it mimics a typical JSON configuration file. We used 100,000 parameters configuration. This is a quick process that results in 100,000 samples that contains some boolean parameters, some values drawn from a normal distribution in which the $mu$ is the same of the observed sample and the $sigma$ is the double of the observed one. Finally, we truncate the distributions to have only positive values in order to guarantee valid parameters. Although we could have a larger variation in the generated parameters, the values need to be both comparable to the observed sample and reasonable within the meaning and mechanisms of the ABM model.  

Then, the trained models are applied to the simulated configuration parameters and predicted results are produced. 
Such a procedure, allows for the investigation of the difference in parameters of the configuration files for those cases that reach the target and those that do not.

\paragraph{Proposal.} We use optimal targets built upon simulation results along with the originating configuration parameters to train models of classification. The trained models are then applied to a large number of random configuration files to predict results. We analyze the configuration parameters relatively to the results they produced. 	

\section{Results and considerations}

We ran tests with a training data of 65\% of the sample and the rest we used as tests \ref{table:1}.\footnote{We also ran tests with normalized data, but the accuracy was lower.} The accuracy of the models was relatively good. However, analysis of the confusion matrix indicated that the goodness of fit came mostly from the non-optimal cases that were correctly predicted. In many cases the number of cases classified as optimal that were optimal were the same as the cases classified as optimal that were not so. \footnote{This was specially the case for other target optimal cases, such as higher GDP and lower GINI.} We believe that what influenced this difficulty of classification is mostly the fact that the sample has - by construction - a very small number of optimal cases. Ideally, we would need to compensate that with a larger number of samples. But that was exactly the motivation of having the ML in the first place.

Having that in mind, we decided to analyze the one case in which the accuracy was very high and the confusion matrix does not have many optimal cases that were incorrectly classified, i.e., we analyzed the Random Forest Classification results.

\begin{table}
\centering
\begin{tabular}{l|cc|cc|cc|cc} 
\toprule
{} & & {Tree} & & {SVC} & & {MPL} & & {Voting} \\
\midrule
{Score} & & 0.9878 & & 0.9634 & & 0.9390 & & 0.9756\\
\midrule
{Confusion} & 75 & 0 & 75 & 0 & 73 & 2 & 75 & 0 \\
{Matrix} & 1 & 6 & 3 & 4  & 3 & 4 & 2 & 5 \\
\bottomrule
\end{tabular}
\label{table:1}
\caption{Results for accuracy and confusion matrix of classification methods tested}
\end{table}

Most of the parameters have the same behavior from the sample and the random generation parameters. That means, most of them have the higher or lower value for the optimal case, as the sample does. 

We consider the two ones that behaved differently of interest, plus the behavior of the boolean rules. 

\begin{enumerate}
\item \textit{Labor\_market} has a 13\% value above the mean for the random generated parameters whereas the value for the sample is 19.8\% below the mean for the optimal case. Labor market is a parameter that determines the frequency that the firm enters the market hiring and firing employees. The ML results suggest that the firm should enter the labor market rather more often than not.
\item \textit{Tax on consumption} for the real case on the optimal case is smaller than the average by a factor of 50\% whereas the random parameter suggests the optimal case would have a value that is 15\% above the average value. This suggests that considering QLI and unemployment for the configuration and mechanisms of the ABM, a higher tax on consumption may benefit the economy. 

\end{enumerate}
Those are differences in terms of direction and magnitude. This may suggest either that the model could be better calibrated or that actual cases would need behavior change in order to move the economies in the direction of the parameters to reach better social results. 

The boolean parameters for the optimal case confirm and helps reinforce the conclusions drawn in \cite{furtado_policyspace:_2018}. That is, that the (i) FPM rule, which is a tax transfer scheme for poorer and smaller municipalities, (ii) that the alternative in which the municipalities are together for fiscal purposes and (iii) that the use by the firm of market information to set wages are all socially beneficial. 

Finally, we can use the score of the cities' parameters in the optimal case as a ranking of the cities. The ranking resulting from the sample cases, has Porto Alegre, Rio de Janeiro, Belo Horizonte, Santos, e Jundia\'{i} as top five ACPs. Interestingly enough (for those who knows the cities), the random optimal case ranking brings other relevant ACPs to the top. It reorders the five first ones and it brings Goi\^{a}nia, Maring\'{a}, S\~{a}o Paulo e Ribeir\~{a}o Preto to the top 9. 

All in all, we used a trained ML to simulate parameters for an optimal case of an ABM model. Results suggest that most of the parameters and the boolean rules are well adjusted with two parameters that should be further investigated. Specifically, the samples of different ACPs was very restricted suggesting that further samples should be provided so that the set of ACPs are better represented in the model.

\paragraph{Notes and Comments.}
This is an initial exploratory exercise delivered as and end-of-course product for a Machine Learning course at Ipea given by Thiago Marzag\~{a}o. We appreciate the help and the classes. 

\bibliography{ml}
\bibliographystyle{acm}
\end{document}